\documentclass{PoS}
\usepackage{amssymb,amsmath,wrapfig}
\usepackage{graphicx} 
\usepackage{dsfont}
\usepackage{bbold}
\usepackage{cancel}
\usepackage{url}
\usepackage{setspace}

\newcommand{\MSbar}{\overline{\text{MS}}}

\graphicspath{{Figures/}}

\title{From spin models to lattice QCD -- \\
the scientific legacy of Peter Hasenfratz}

\ShortTitle{From spin models to lattice QCD}

\author{\speaker{Urs Wenger} \\
        Albert Einstein Center for Fundamental Physics\\
         Institute for Theoretical Physics\\
        University of Bern\\
        Sidlerstrasse 5\\
        CH--3012 Bern\\
        Switzerland\\
        E-mail:  \email{wenger@itp.unibe.ch}}

      \abstract{This is a transcript of my conference talk in
        remembrance of Peter Hasenfratz who deceased earlier in
        2016. One of Peter's many important contributions to the
        lattice community has been the initiation of the first lattice
        conference at CERN in 1982. From among his many important ideas,
        which Peter contributed to our field, I choose to discuss three in some
        detail and show how they are influencing the subject today.
      }

\FullConference{34th annual International Symposium on Lattice Field
  Theory\\
 24-30 July 2016\\
University of Southampton, UK}

\begin{document}
\begin{wrapfigure}{l}{0.4\textwidth} 
\centering
\includegraphics[width=0.4\textwidth]{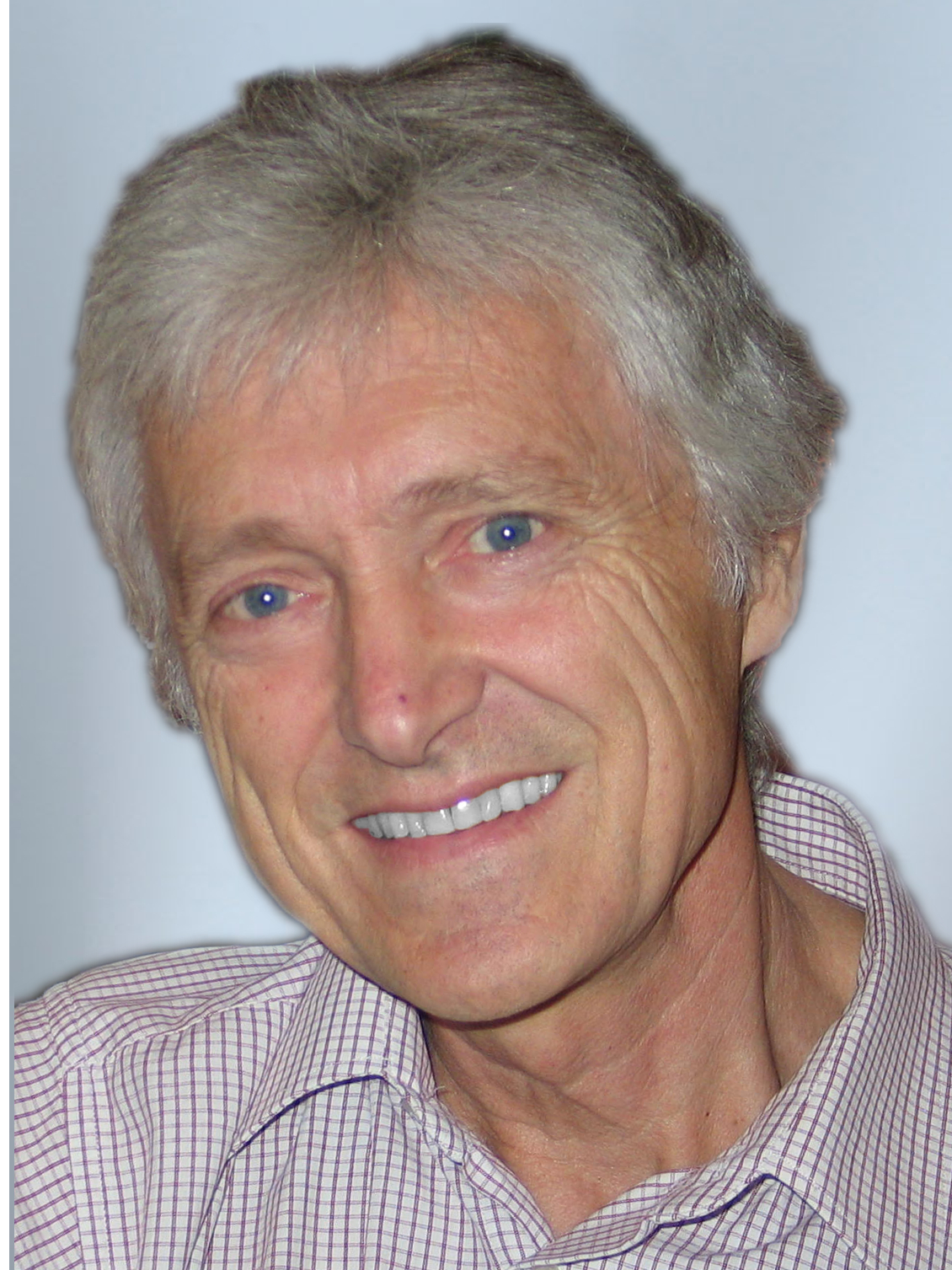}\\
\vspace{0.25cm}
P\'eter Hasenfratz 1946 -- 2016
\end{wrapfigure}
It is with great sorrow that we have to accept the fact that Peter
Hasenfratz deceased earlier this year on April 9, 2016 at the age of
69 as a consequence of his suffering from Alzheimer's disease. Our
condolences go out to his family, in particular also to his sister and
our colleague Anna, and to his friends and colleagues.

Peter was a very gentle and extremely modest person and he had a very
generous personality. His generosity did not only reach out to his
family, but also to his friends and colleagues. His kindness is very
nicely captured by a quote of an old friend of Peter:
``{\it If I would not know P\'eter's profession, I would have guessed
  he must be a pediatrician}". Peter's softness is also reflected in a
statement by Peter himself, which he once mentioned to me: ``{\it If I
  would not have become a scientist, I would be a poet}''.  I think
this quote expresses quite well his personality, but also his special
way how he looked at nature. Nevertheless, despite his poetic approach
to physics, he would tolerate only the highest scientific standards
and was very tough when it came to discussing physics.
--
We are sad to have lost an extraordinary person and scientist.

Before appreciating some of Peter's scientific achievements, let me
briefly recall his CV. Peter was born in Budapest on September 22,
1946. After his B.Sc.~in Physics in 1971 at E\"otv\"os University,
Budapest, he continued studying physics and received his Ph.D.~degree
in 1973, also from E\"otv\"os University. Subsequently, he became a
member of the Central Research Institute for Physics (CRIP) in
Budapest, before taking a postdoc position from 1975 to 1976 in
Utrecht, working together with Gerard 't Hooft. After two years back
at CRIP he became a postdoc at CERN in 1979 and was promoted to a
staff member at the Theory Division in 1981. Only three years later he
got an associate professorship at the Institute for Theoretical
Physics (ITP) of the University of Bern in 1984 and was promoted to
full professor in 1991. With intermittent sabbatical stays in Santa
Barbara, Boulder, Boston and CERN he stayed at the ITP until 2011 when
he retired.

\section{Peter Hasenfratz' scientific contributions}
It is difficult to summarize Peter's vast scientific contributions in
just a short amount of time, and it is even more difficult to
acknowledge it in terms of numbers.  Peter has over 125 published
articles which by today's standards does not seem a lot. I think this
simply reflects the fact that his publications only followed very high
quality standards which were difficult to satisfy, not least by
himself. He would publish only when he considered the results
worthwhile and significant. Many of the publications are in fact
proceedings of plenary talks, review articles, and lecture notes. Not
suprisingly, he was an excellent academic teacher, not only during his
lectures at the university, but also at numerous international
schools. As a student it was difficult not to get infected by his
enthusiasm about quantum field theory (QFT) in particular and particle
physics in general. Consequently enough, he has also been an advisor
to a total 14 Ph.D.~students.

One particular scientific contribution is actually of sociological
nature. In 1982 Peter initiated the first lattice conference as a
workshop at CERN. In the unimaginable times without internet, he
issued handwritten personal invitations to all participants. In this
way, he put the seed for a new scientific community, which over the
years has grown into what we experience today at this conference. So
far, there have been 33 conferences with several hundreds of
participants each - the 34th symposium here in Southampton for example
boasts more than 420 participants.

\begin{wrapfigure}[21]{r}{0.4\textwidth} 
{\centering
\includegraphics[width=0.4\textwidth]{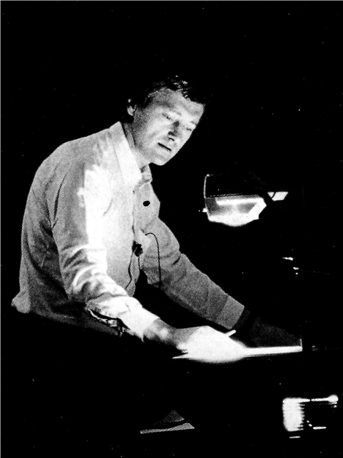}
}
\vspace{-0.55cm} 
\caption { Peter Hasenfratz delivering a plenary talk on the mass
  limits of the Higgs particle at the lattice conference at Fermilab
  in 1988 \cite{Hasenfratz:1988ts} (picture from
  \cite{Hasenfratz:1988cc}).}
\end{wrapfigure}
Peter seemed to have an ability to analytically calculate seemingly
uncalculable things. Let me here just mention two specific examples
which I will discuss in more detail later, the scale parameter of QCD
on the lattice on the one hand, and the exact mass gaps in several
two-dimensional asymptotically free quantum field theories on the
other hand. He was a very creative and original thinker and provided
numerous seminal contributions to lattice QFT, such as the concept of
Fixed Point (FP) actions, the index theorem, the understanding of
lattice chiral symmetry, the proper lattice definition of the chemical
potential, and so on and on, just to mention here the concepts which I
will refer to later in my talk. Other major research topics to which
Peter contributed include the quark bag model, topological
excitations, spin models, hopping expansion, Higgs physics (upper
bound, top quark condensate, \ldots), finite size effects from
Goldstone bosons, or finite temperature phase transitions in QCD.
Higgs physics was a topic Peter was particularly fond of. The picture
on the right shows him delivering a plenary talk on the mass limits of
the Higgs particle at the lattice conference at Fermilab in
1988. Throughout his career he continued to be fascinated by the Higgs
mechanism and the underlying field theoretic concepts, and he
certainly enjoyed the discovery of the Higgs particle a few years ago.

It would be impossible to cover all of the topics mentioned above -
instead I will just focus on three separate topics which I find
particularly interesting, not least because they connect to issues
which are going to be discussed at this conference. The first topic
concerns the connection between the lattice and the continuum, the
second concerns the calculation of mass gaps, and the third the
renormalization group (RG) and fixed point (FP) actions.

\section{The connection between the lattice and the continuum}
Let us turn back the clock about 35 years into a time when lattice
field theory, and in particular numerical calculations, started to
come out of its infancy. Peter was very busy as a salesman in the
high-energy particle physics community, giving many talks on the
lattice approach to field theory in general, and QCD in
particular. Despite being a salesman, he did never stop to emphasize
the shortcomings of these early calculations, to many of which Peter
contributed himself, and he kept being very critical towards them.
Let me quote in the following a few statements which can be found in
his review paper on lattice gauge theories from 1981 \cite{Hasenfratz:1981ua}: 
"{\it MC simulations did
  not help us to obtain a better physical understanding, a deeper
  insight into the theory.}" -- 
"{\it Is $g$ small enough? [\ldots] there is reason to worry: the
  approach to asymptotic scaling might be very slow.}" --
"{\it In spite of the intense work, there is no real progress one can
  report on.}" -- 
"{\it The whole program is faced with unexpected and unpleasant
  difficulties at this moment.}" -- 
"{\it Clarification is needed.}" --
"{\it One should consider these numbers with some reservations.}" --
"{\it [\ldots] although it is not clear whether every part of the
  calculation is under control.}" --  
%\noindent
Further into 1982 \cite{Hasenfratz:1982sa} he continued along the same
line, always keeping expectations low and advocating a slow and
careful approach, instead of a fast and attention-grabbing one:
"{\it [\ldots] the reliability of this procedure is really
  questionable.}" --
"{\it I sense a big change concerning the expectations of the physics
  community. Actually I believe this change is too big.}" --
"{\it Please, have your own healthy doubts 
  [\ldots]. Solving QCD is not \underline{so} easy.}" --
"{\it [\ldots] admit clearly the defects of our methods and make
  serious efforts to improve them. This path is less spectacular, but,
  perhaps, worth following.}" --
All these statements emphasize the carefulness and great
diligence which Peter applied to the lattice approach, and I would
like you to keep this in mind, when I discuss some of the very latest
results from lattice QCD at the end of this section.

In the late seventies and early eighties it was not yet clear whether
lattice QCD provides a sensible regularization of continuum
QCD. Hence, in order for lattice QCD to be meaning- and useful on a
practical level, it was obviously crucial to establish the connection
between calculations on the lattice and in the continuum. Lattice QCD,
either in the chiral or the quenched limit, contains only the
dimensionless coupling $g$ and implicitly the dimensionful lattice
spacing $a$ as parameters. For a physical mass $m$, or equivalently a
correlation length $\xi$, one therefore has
\[
m = f(g) \cdot \frac{1}{a}  \quad \text{or} \quad \xi = h(g)\cdot a \, .
\]
The continuum limit is reached when $1/m$ or $\xi \gg a$, i.e.~when
the lattice system approaches a continuous phase transition.  In
asymptotically free theories, the limit is straightforward to reach,
since the lattice spacing $a \rightarrow 0$ for $g \rightarrow 0$.  By
changing $a$ and $g$ appropriately towards the continuum limit,
physical quantities should become independent of $a$,
\[
\frac{d}{da} m = 0 \quad (a \rightarrow 0)\, ,
\]
which is equivalent to say that the theory is renormalizable. The
renormalizability uniquely fixes the relation between $a$ and $g$
through the differential equation for $f(g)$,
\begin{equation}
-f(g) + f'(g) \left(a \frac{d}{da} g \right) = 0 
\qquad \text{where} \qquad\beta(g) \equiv a \frac{d}{da} g = -b_0 g^3 - b_1 g^5 - \ldots \, .
\label{eq:beta-function}
\end{equation}
 Hence, every physical quantity on the lattice can be
expressed in terms of a {\it single}, RG-invariant mass parameter
$\Lambda^\text{latt}$, e.g. $m = c_m \cdot \Lambda^\text{latt}$, and
the dependence of the scale on the gauge coupling $g$ is determined by
Eq.~(\ref{eq:beta-function}), yielding
\[
\Lambda^\text{latt} = \frac{1}{a}\, e^{-1/2b_0 g^2} \left(b_0
g^2\right)^{-b_1/2b_0^2} \, \cdot [1 + {\cal O}(g^2)]
\]
to lowest order in perturbation theory. Analogously, in a continuum
renormalization scheme one has
\[
\Lambda = M \, e^{-1/2b_0 g(M)^2} \left(b_0
g(M)^2\right)^{-b_1/2b_0^2} \, \cdot [1 + {\cal O}(g(M)^2)]
\]
where the mass parameter $M$ corresponds to the renormalization scale
introduced in the continuum renormalization scheme.

So, in order to
set the scale in the lattice theory, and to make sense of it, one
better connects the two scales. In 1980 Anna and Peter Hasenfratz
calculated this connection \cite{Hasenfratz:1980kn}:
\begin{equation*}
\Lambda^\text{MOM}_\text{\tiny Feynman gauge} = 83.5
\, \Lambda^\text{latt} \quad \text{for} \,\, SU(3), \quad \quad
\Lambda^\text{MOM}_\text{\tiny Feynman gauge} = 57.5 \, \Lambda^\text{latt} \quad \text{for} \,\, SU(2) \, .
\end{equation*}
The computation involves a rather long and tedious 1-loop calculation
of 2- and 3-point functions in lattice perturbation theory. In
particular, the calculation provides an explicit demonstration that
there are no unwanted divergences and that all noncovariant terms
cancel. In that sense, the result truly constitutes a milestone in
establishing lattice QCD as a viable and useful regularization of
QCD. Anna and Peter were the first to get the relation correct,
thereby settling a dispute which was ongoing at the time. It is hard
to overestimate the challenge and the difficulty of this calculation,
and in fact it took 15 more years until the corresponding 2-loop
calculation was completed \cite{Luscher:1995np}.

Of course,  the $\Lambda$ parameter is nonperturbatively defined,
\begin{equation*}
\Lambda = M \, e^{-1/2b_0 g(M)^2} \left(b_0
g(M)^2\right)^{-b_1/2b_0^2} 
\times\exp\left[- \int_0^{g(M)} dx\left(\frac{1}{\beta(x)} +
      \frac{1}{b_0 x^3} - \frac{b_1}{b_0^2x}\right) \right] \, ,
\end{equation*}
and lattice QCD is the ideal method to relate it {\it
  nonperturbatively} to the low-energy properties of
\begin{wrapfigure}[16]{r}{0.4\textwidth}
\centering
\vspace{-0.3cm}
%\hspace{-0.5cm}
\includegraphics[width=0.4\textwidth]{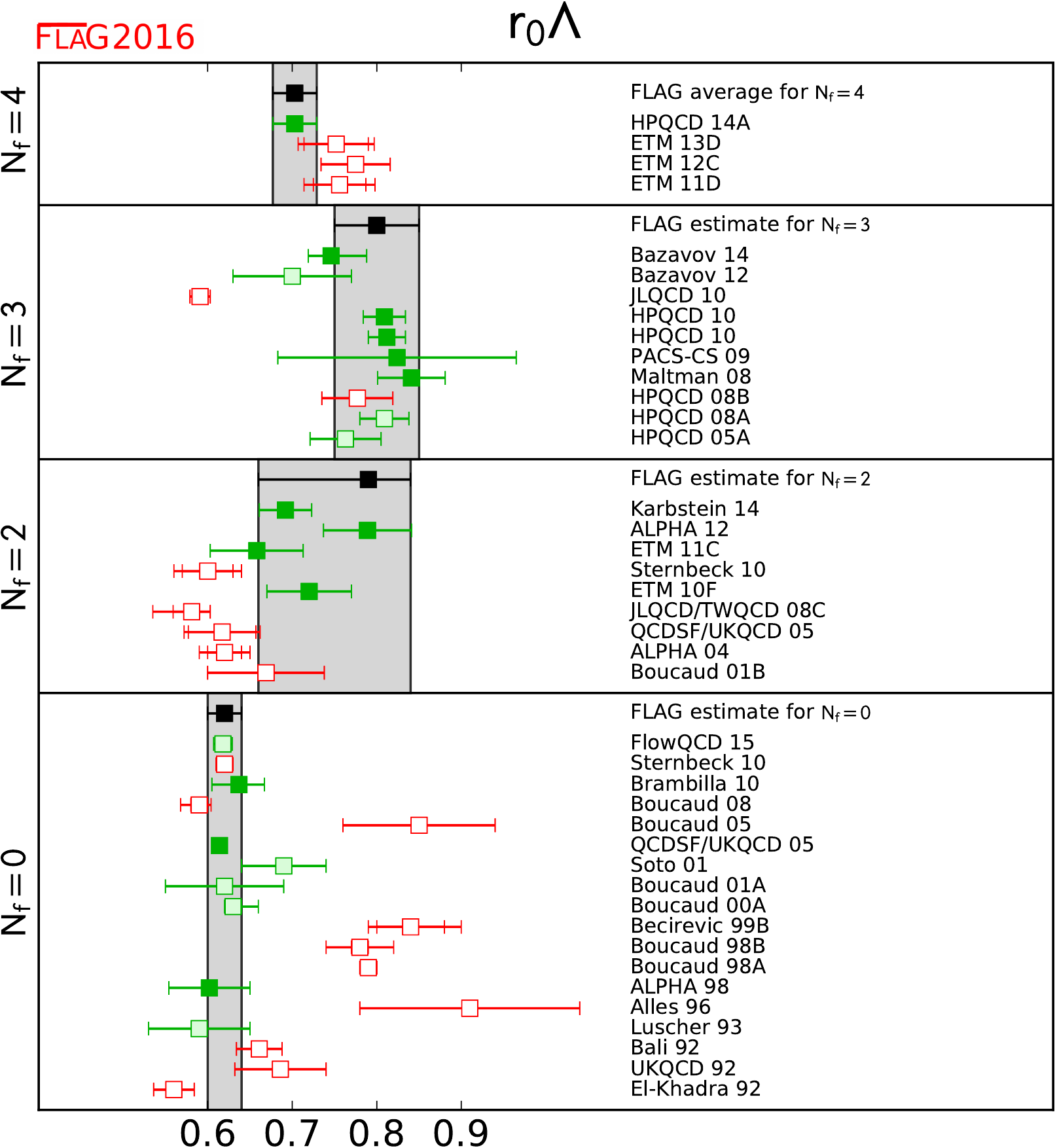}
%\vspace{-0.25cm}
\caption{Lattice QCD results for the
$\Lambda$ parameter in the $\MSbar$-scheme in units of $r_0$ from the
FLAG report \cite{Aoki:2016frl}.
\label{fig:Lambda-parameter}}
\end{wrapfigure}
QCD. In fact, the $\Lambda$ parameter is a quantity which nowadays is
rather well determined from lattice QCD calculations. In
Fig.~\ref{fig:Lambda-parameter} we show the latest collection of
lattice QCD results for the $\Lambda$ parameter in the $\MSbar$-scheme
in units of $r_0$ from the FLAG report \cite{Aoki:2016frl}. Obviously,
while the results for $N_f=0$ and 2 flavour QCD are of no interest
from a phenomenological point of view, they certainly are for the
theoretical understanding of QCD in general.
\begin{figure}[t]
\centering
\vspace{-0.5cm}
\includegraphics[width=0.45\textwidth]{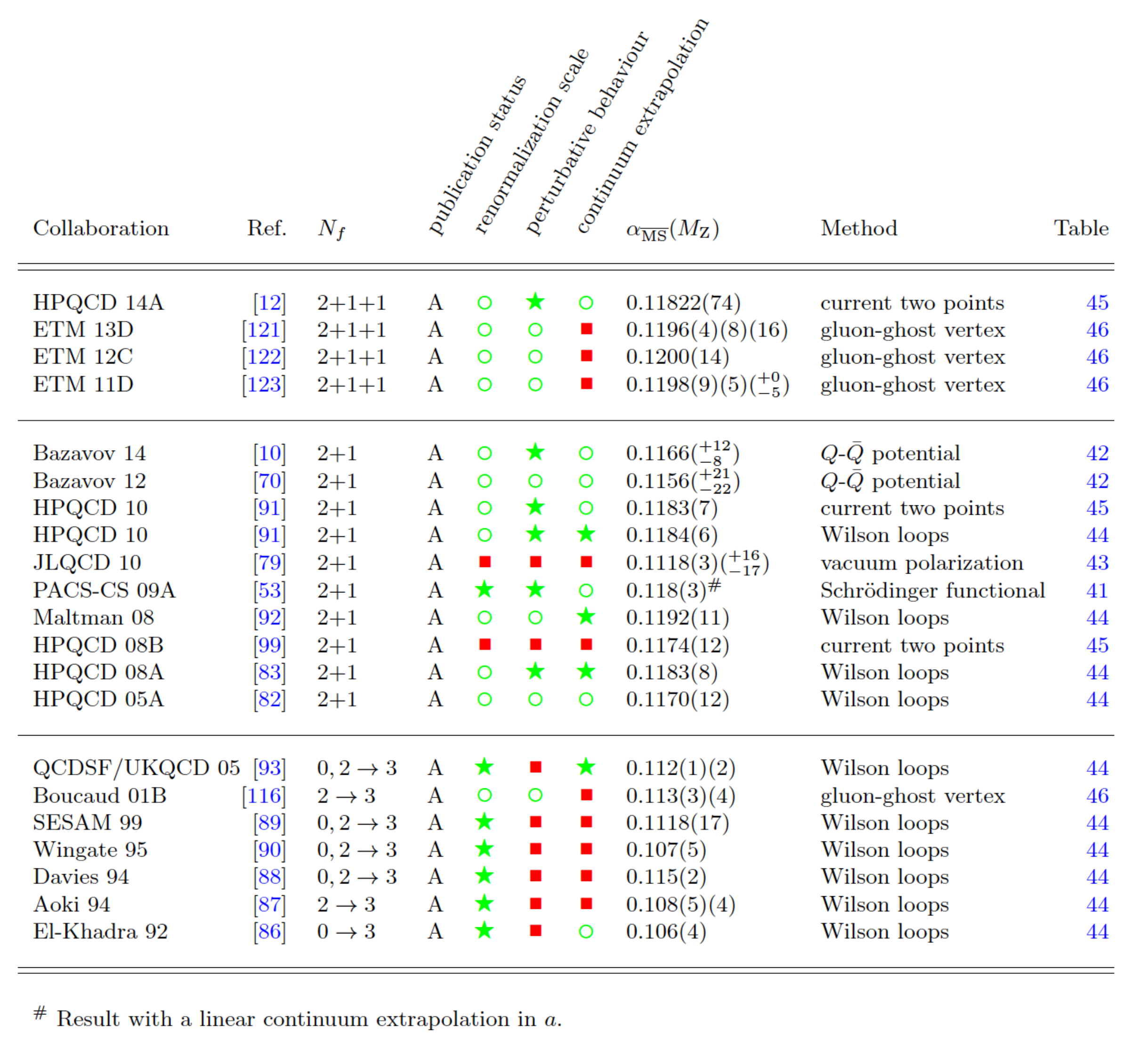}
\hfill
\includegraphics[height=6.0cm]{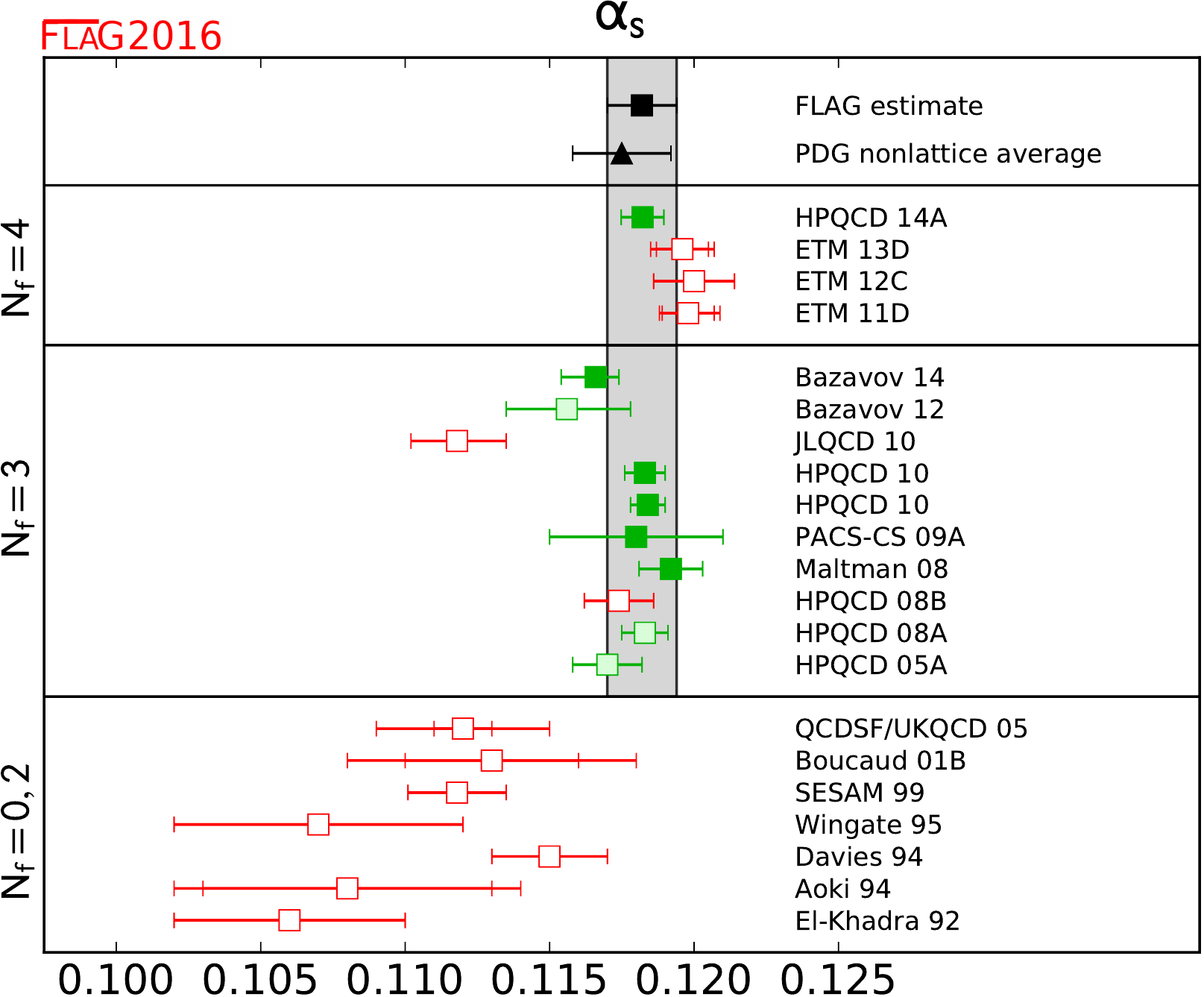}
\caption{Collection of lattice QCD results for the  $N_f=5$  strong coupling $\alpha_{\MSbar}^{(5)}(M_Z)$ in the
$\MSbar$-scheme at the scale $M_Z$ from the
FLAG report \cite{Aoki:2016frl}.\label{fig:alpha_s}}
\vspace{-0.5cm}
\end{figure}

Closely related to the $\Lambda$ parameter is the running strong
coupling $\alpha_s$ at the renormalization scale $M$,
\[
\alpha_s(M) = \frac{g^2(M)}{4 \pi} \, .
\]
It can for example be determined by measuring a short distance
quantity ${\cal O}$ at scale $M$ and matching it with the
corresponding perturbative expansion in terms of the coupling in the
$\MSbar$-scheme,
\[
{\cal O}(M) = c_1 \alpha_{\MSbar}(M) + c_2 \alpha_{\MSbar}(M)^2 +
\ldots \, .
\]
Also for this quantity, lattice QCD calculations are well advanced, as
can be seen from the table and plot in Fig.~\ref{fig:alpha_s}.  Many
collaborations provide values for the $N_f=5$ strong coupling
$\alpha_{\MSbar}^{(5)}(M_Z)$ in
\begin{wrapfigure}[21]{r}{0.3\textwidth} 
\includegraphics[width=0.3\textwidth]{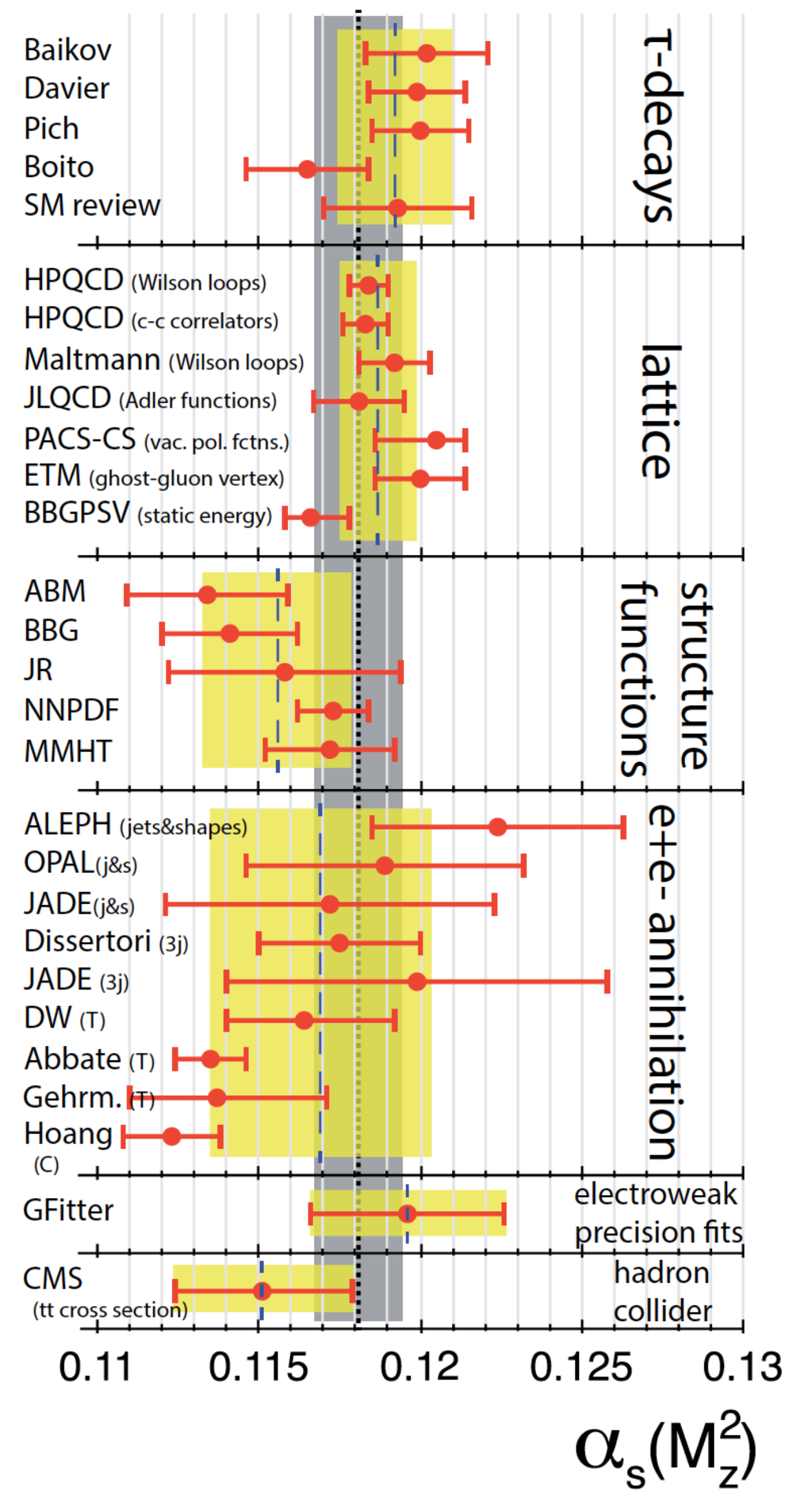}
\caption{Summary of determinations of $\alpha_{\MSbar}(M_Z^2)$ from
 the 2016 edition of the  PDG \cite{Olive:2016xmw}.\label{fig:alphas_PDG}}
\end{wrapfigure} 

\noindent
the $\MSbar$-scheme at the scale given by the $Z$-boson mass $M_Z$,
employing a variety of different short distance quantities. It is
clear, that a critical assessment of the situation is necessary in
order to provide a consistent picture and a reliable estimate of the
strong coupling useful for phenomenology. This is exactly what FLAG
provides in its review \cite{Aoki:2016frl}.
A careful evaluation indicates that for the strong coupling the
dominant source of uncertainty comes from discretization errors and
the truncation of continuum and lattice perturbation theory.

It is interesting to compare the FLAG 16 lattice average
$\alpha_{\MSbar}^{(5)}(M_Z) = 0.1182(12)$ with the values from the
2016 edition of the PDG \cite{Olive:2016xmw}.  They quote
$\alpha_{\MSbar}^{(5)}(M_Z) = 0.1174(16)$ for an average of all
nonlattice results, and $\alpha_{\MSbar}^{(5)}(M_Z) = 0.1181(11)$ for
the world average including the results from the lattice. It is clear
from Fig.~\ref{fig:alphas_PDG} that the lattice determination by now
provides the most precise value. It is gratifying to see that after
more than three decades the common effort of the lattice community
finally starts to pay off.

\begin{figure}[!t]
\centering
\includegraphics[width=0.49\textwidth]{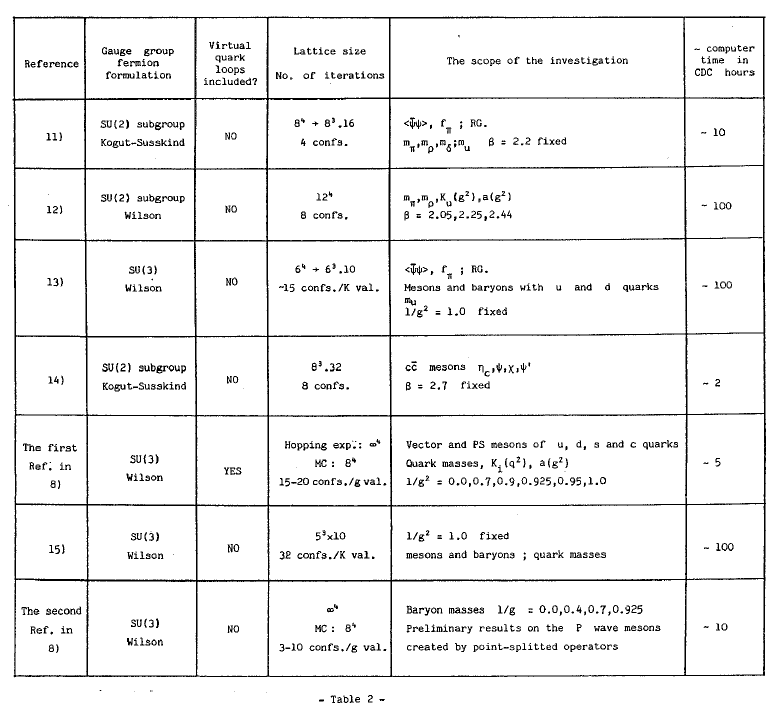}
\hfill
\includegraphics[width=0.45\textwidth]{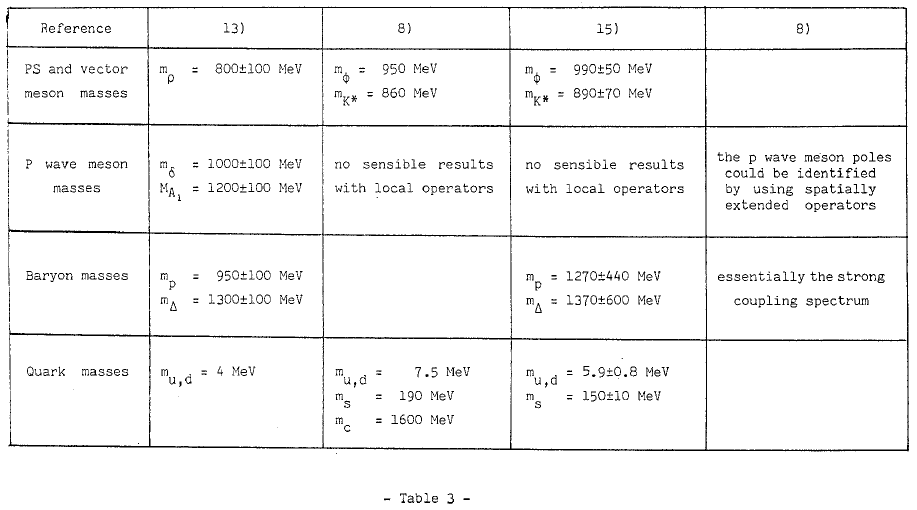}
\caption{Early tables for critically assessing and summarizing lattice
  calculations compiled by Peter in 1982 \cite{Hasenfratz:1982sa}.\label{fig:old_tables}}
\end{figure}

In fact, as compared to its 2014 edition, the PDG is now following a
more conservative approach for averaging the lattice results, very
much in the spirit of FLAG 16, and even more so in the spirit of
Peter. Indeed, it seems that Peter was right 35 years ago, when he
advocated to follow a conservative approach, even though it is less
spectacular.
The conservative approach followed by FLAG requires a critical
assessment of the available lattice results, as summarized e.g.~in
Fig.~\ref{fig:alpha_s}. However, such tables and summaries are not a new
invention. In fact, back in 1982 Peter has already provided such
tables, cf.~Fig.~\ref{fig:old_tables}, summarizing the properties of
various lattice calculations and the corresponding results available
at that time \cite{Hasenfratz:1982sa}.  Despite being very critical,
Peter was nevertheless always very optimistic, as the following quote
by Peter in  \cite{Hasenfratz:1981ua} nicely portrays:
"{\it We are able to obtain non-perturbative numbers in a
  four-dimensional, relativistic, relevant theory. We are proud
    of it.}"

\section{The mass gaps}

One of the most difficult problems in a quantum field theory is the
determination of the relation between the renormalized couplings from
the Lagrangian and the physical masses of the theory, such as for
example the nucleon mass in the chiral limit of QCD in units of
$\Lambda^{\MSbar}$,
\[
m_N = c_{m_N} \cdot \Lambda^{\MSbar} \, .
\]
The difficulty lies in the fact that the Lagrangian is defined at
short distances (UV-scale), while the masses are parameters at large
distances (IR-scale). Surprisingly, there is a family of models where
this relation can be found {\it exactly}, namely the O($N$) nonlinear
sigma models in $d=2$ dimensions. For $N\geq 3$ these integrable
models are asymptotically free and contain massive O($N$) isovector
multiplets. In 1990 Peter, together with Michele Maggiore and Ferenc
Niedermayer, calculated this relation {\it exactly} for $N=3$ and 4
\cite{Hasenfratz:1990zz}:
\begin{align*}
m &= \frac{8}{e} \cdot \Lambda^{\MSbar} \hspace{2.1cm} N=3 \, ,\\
m &= 
\sqrt{\frac{32}{\pi e}} \cdot\Lambda^{\MSbar} \hspace{1.5cm} N=4 \, .
\end{align*}
In the same year, Peter and Ferenc extended the calculation to
arbitrary $N \ge 3$ \cite{Hasenfratz:1990ab}:
\[
m = \left(\frac{8}{e}\right)^{1/(N-2)} \frac{1}{\Gamma(1+1/(N-2))}
\cdot\Lambda^{\MSbar} \, .
\]
It is interesting to note that at the time, there were over 30
nonperturbative determinations which differed wildly from each other.

While the calculation is rather involved, it is based on a beautiful
idea. It starts from the introduction of a chemical potential $h$
coupled to a Noether charge and the observation that the {\it change}
of the free energy is RG invariant, as is the chemical potential $h$
itself. Then, on the one hand, the free energy can be calculated in
perturbation theory in the regime $h \gg m$ where the theory becomes
asymptotically free,
\begin{equation*}
f(h) - f(0) = 
-(N-2)\frac{h^2}{4\pi} \left[\ln \frac{h}{e^{1/2}
    \Lambda_{\MSbar}} 
+\frac{1}{N-2}\ln \ln \frac{h}{\Lambda_{\MSbar}}
    + {\cal O}\left(\frac{\ln
        \ln(h/\Lambda_{\MSbar})}{\ln(h/\Lambda_{\MSbar})}
    \right)\right] \, .
\end{equation*}
On the other hand, since the model is integrable, the free energy can
also be calculated by applying the Bethe ansatz, or directly from the
$S$-matrix,
\[
f(h) - f(0) = -\frac{m}{2\pi} \int \cosh\theta \, \varepsilon(\theta) d\theta
\]
with $\varepsilon(\theta)$ fulfilling a specific integral equation
\cite{Hasenfratz:1990ab}. One can then use a generalized Wiener-Hopf
technique to express the integral equation in terms of $\ln h/m$,
again in the regime $h\gg m$, and read off the mass $m$ in terms of
$\Lambda_{\MSbar}$ by comparing the expressions obtained from the
two approaches. The same idea can be applied to other quantum field
theories. Its application yields for example also the exact mass gap
in the Gross-Neveu model \cite{Forgacs:1991rs} and in the $d=2+1$
dimensional antiferromagnetic Heisenberg model at low temperatures
\cite{Hasenfratz:1990jw,Hasenfratz:2005fn}.

The reason why I dwell on this in some detail is due to the fact that
the idea to couple a chemical potential to a conserved charge and to
calculate the corresponding change in the free energy is very
general. In fact, it is relevant and in use even today. One
interesting, very recent application is for matching chiral
Lagrangians for QCD with different regularizations
\cite{Niedermayer:2016yll,Niedermayer:2016ilf}. The calculation is
related to the QCD rotator in the $\delta$-regime where $m_\pi L_s \ll
1$ and $F_\pi L_s \gg 1$~\cite{Leutwyler:1987ak} and provides a
promising new way to determine the low-energy constants of QCD
\cite{Hasenfratz:2009mp} by introducing an infrared cutoff through a
finite spatial box size $L_s$ and then studying the finite size
scaling of the spectrum in the chiral limit. More precisely, the
chiral Lagrangian for massless 2-flavour QCD has a $\text{SU}(2)
\times \text{SU}(2) \simeq \text{O}(4)$ symmetry and the general
O($N$) spectrum is given by a quantum mechanical rotator
\begin{equation*}
E(l) = l(l+N-2)/2 \Theta \, \qquad l =0,1,2,\ldots
\end{equation*}
where $\Theta = F^2L_s^3$ is the moment of inertia of the
rotator~\cite{Leutwyler:1987ak}. The next-to-leading (NLO) term of the
expansion in $1/(F^2L_s^2)$ has been calculated by Peter and Ferenc
some time ago \cite{Hasenfratz:1993vf}, while the NNLO terms were
obtained more recently in the dimensional regularization (DR) scheme
\cite{Hasenfratz:2009mp} and on the lattice
\cite{Niedermayer:2010mx}. Finally, the very recent, heroic
calculation by Niedermayer and Weisz
\cite{Niedermayer:2016yll,Niedermayer:2016ilf} makes use of the same
idea and thereby connects the couplings in the two regularizations of
the effective theory, i.e.~it converts the expressions for physical
quantities obtained in the lattice scheme to the corresponding ones in
the DR scheme. In particular, it relates the finite-volume mass gap on
the lattice to the DR scheme.

\section{The renormalization group and FP actions} 
\begin{figure}[htb]
\centering
\includegraphics[angle=-0.75,width=0.5\textwidth]{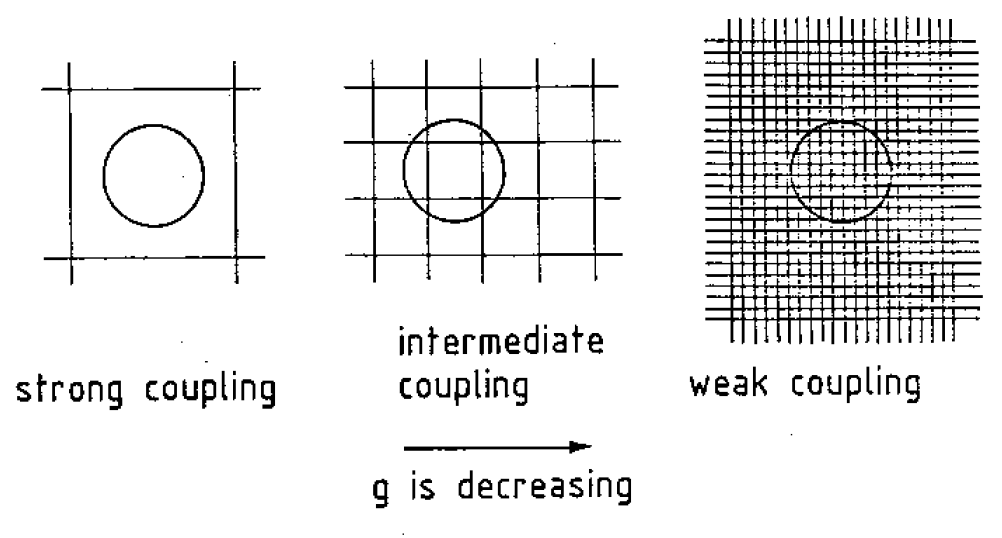}
\caption{The continuum limit as the critical limit of a statistical
  lattice system where the lattice spacing becomes small compared to
  the characteristic physical distances. Figure from
  \cite{Hasenfratz:1983vk}. \label{fig:critical limit}}
\end{figure}

Peter had a very deep appreciation and understanding of the Wilson
renormalization group (RG) description of quantum field theories. In
this context, he always emphasized the viewpoint that lattice gauge
theory is just a statistical system, albeit a rather unusual one due
to the local gauge invariance, and that the critical limit
corresponds to the continuum limit of the corresponding quantum field
theory, as illustrated in Fig.~\ref{fig:critical limit} from one of Peter's
lecture notes \cite{Hasenfratz:1983vk}.
Since the lattice provides a fully nonperturbative description of the
phase transition, the continuum physics is recovered in the lattice
system at long distances. Close to the phase transition one can
integrate out the variables describing the short distance lattice
physics and obtain an effective action for the relevant long distance
variables in terms of the effective couplings $\left\{K_\alpha
\right\}$,
\[
\left\{K_\alpha^{(1)} \right\}
\quad       \stackrel{\text{RG}}{\longrightarrow} \quad
\left\{K_\alpha^{(2)} \right\}
\quad       \stackrel{\text{RG}}{\longrightarrow} \quad
\cdots
\quad
\stackrel{\text{RG}}{\longrightarrow}
\quad
\left\{K_\alpha^{(n)} \right\}
\quad   \stackrel{\text{RG}}{\longrightarrow} \quad
\cdots \, .
\]
The sequence of RG transformations might have a fixed point (FP),
\[
\left\{K_\alpha^{*} \right\} \quad
\stackrel{\text{RG}}{\longrightarrow} \quad \left\{K_\alpha^{*}
\right\} \, .
\]
In particular, one is interested in a FP where the correlation length
of the system is $\xi=\infty$. For gauge theories,
\begin{figure}[htb]
\centering
\includegraphics[width=0.4\textwidth]{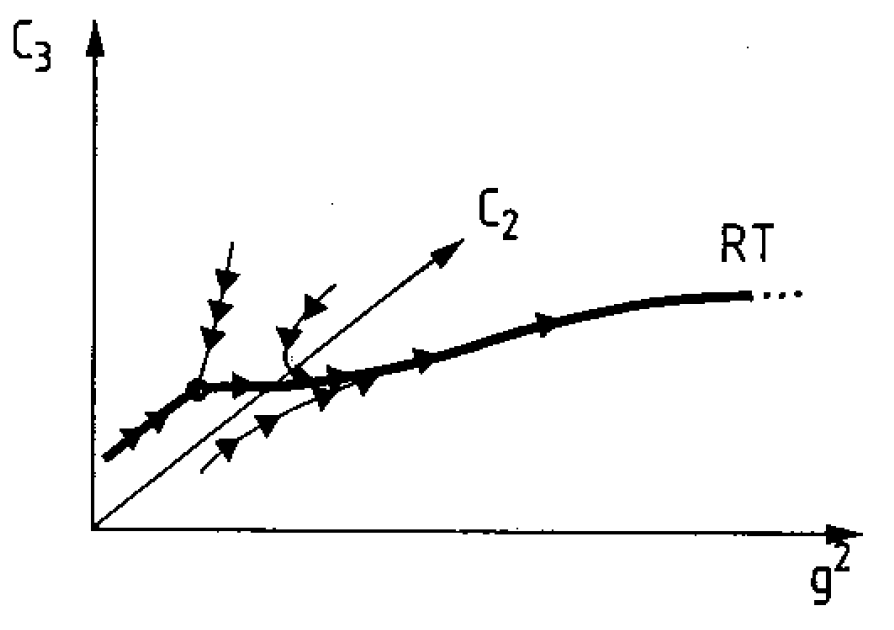}
\caption{Renormalized trajectory (RT) in the parameter space of gauge
  couplings. It comes out of the critical hyperplane $g = 0$ and
  attracts all the flow lines starting in the neighbourhood of the
  fixed point. Figure from
  \cite{Hasenfratz:1983vk}. \label{fig:RG_flow_AF}}
\end{figure}
the RG transformations are complicated due to the requirement of gauge
invariance, but once this is fulfilled, the RG transformations provide
the basic starting point in expecting {\it renormalizability} and {\it
  universality} along the renormalized trajectory (RT) of the lattice
system, cf.~Fig.~\ref{fig:RG_flow_AF} taken from
\cite{Hasenfratz:1983vk}.

The fact that the lattice provides a fully nonperturbative description
of the RG flow of the couplings and the corresponding FP structure
becomes again important today for investigations of quantum field
theories beyond the Standard Model (BSM), as also discussed
extensively at this lattice conference.  It should also serve as a
warning for such investigations. Some of the BSM theories are expected
to possess a conformal FP, however, an IR FP is in general not
perturbative and perturbative intuition could therefore well be
misleading. I think it is important to keep this in mind when
interpreting some of the results from the numerical BSM calculations
dicsussed at this conference, cf.~the plenary talk
\cite{Pica:2017gcb}.

\begin{figure}[bth]
\centering
\begin{minipage}{0.4\textwidth}
\includegraphics[width=\textwidth]{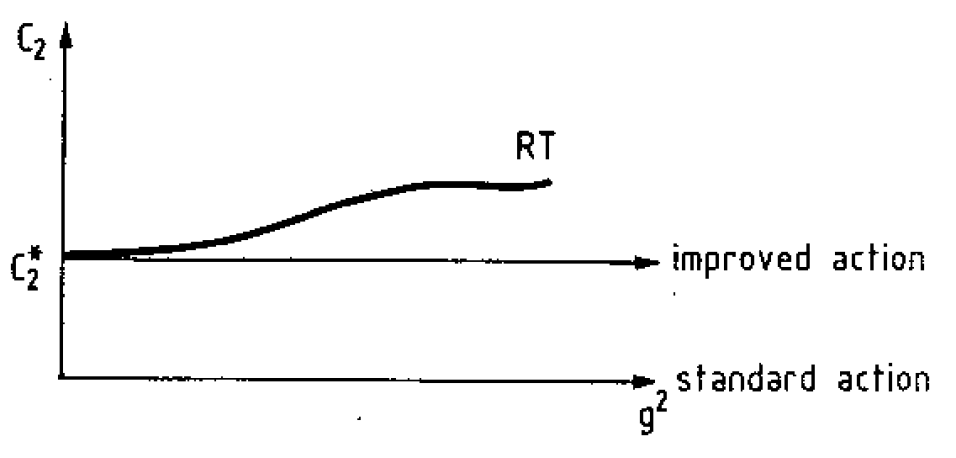}
\end{minipage}
\hfill
\begin{minipage}{0.5\textwidth}
%\hfill 
\includegraphics[width=\textwidth]{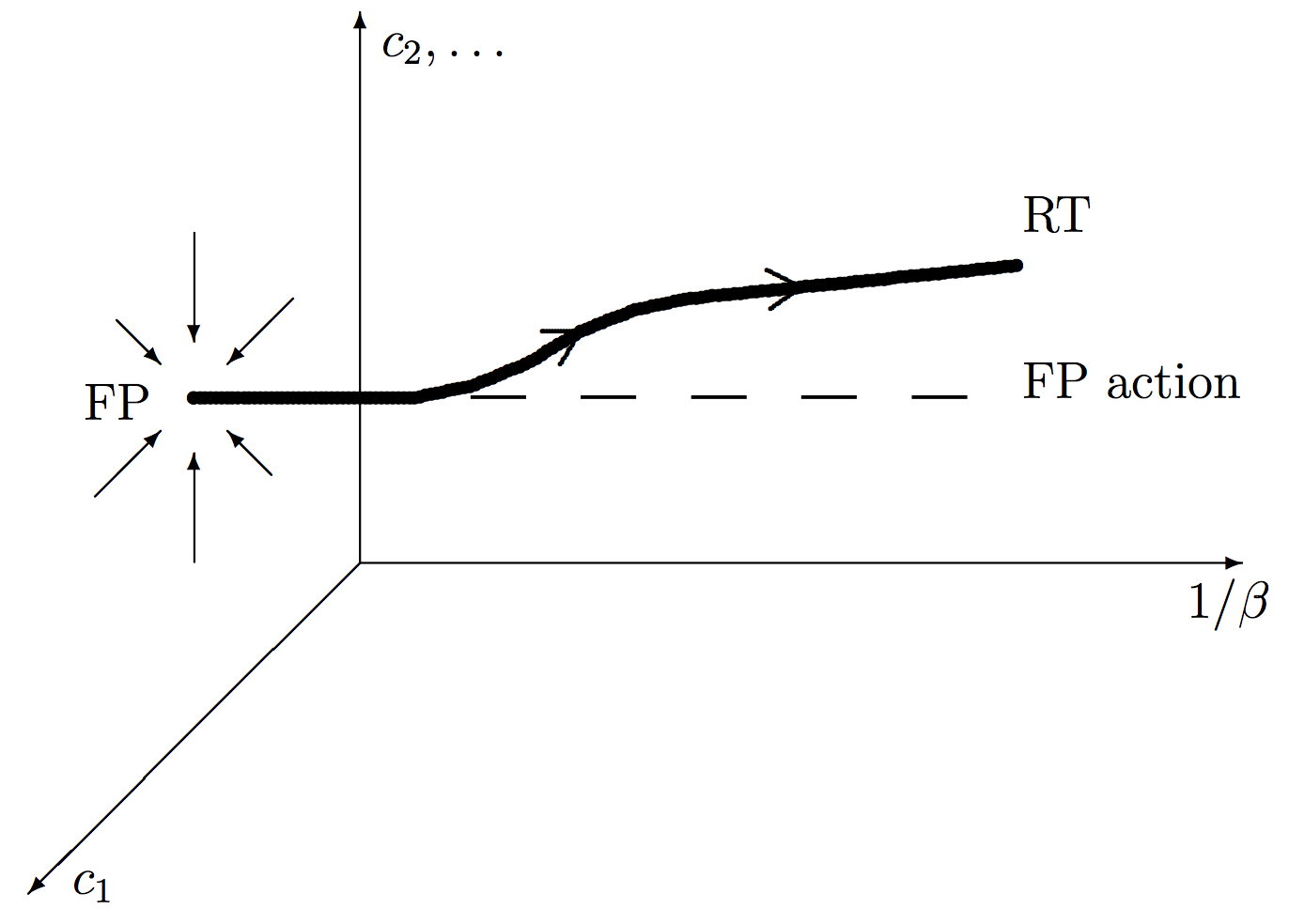}
\end{minipage}
\caption{Renormalized trajectory in the parameter space of (gauge)
  couplings and the improved actions based on an approximation of the
  FP (left plot) or the exact FP (right plot). The
  left plot is from 1983 \cite{Hasenfratz:1983vk}, the right plot from
  1993 \cite{Hasenfratz:1993sp}.
\label{fig:RG_flow_FP}}
\end{figure}

Back in the early eighties, the application of RG ideas to lattice
gauge theory inspired the investigation of perturbative and
nonperturbative improvement of lattice actions by making use of
approximations to the FP, as e.g.~illustrated in the left plot of
Fig.~\ref{fig:RG_flow_FP} taken from Peter's lecture notes from 1983
\cite{Hasenfratz:1983vk}. While the concept of "(quantum) perfect"
actions was already known then, it would take 10 years to make it into
something more specific. In 1993 Peter and Ferenc Niedermayer realized
that for asymptotically free theories the path integral defining the
FP for RG transformations reduces to {\it classical saddle point
  equations},
\[
S_G^\text{FP}[V] = \min_{\{U\}}\left[S_G^\text{FP}[U] +
  T_G[V,U]\right] \, 
\]
where $S_G^\text{FP}$ is the FP gauge action, $T_G$ a blocking kernel
defining a RG transformation, while $U, V$ are the gauge fields on the
fine and coarse lattices, respectively, related by the RG
transformation~\cite{Hasenfratz:1993sp}. The FP Dirac operator is
similarly defined by
\[
D^\text{FP}[V]^{-1} = R[V] + \omega[U]\cdot  D^\text{FP}[U]^{-1} \cdot\omega[U]
\]
where $\omega$ defines a blocking kernel for the fermion fields. It
can then be shown that the action $\beta S_G^\text{FP} +
\overline{\psi} D^\text{FP} \psi$ is {\it classically perfect},
i.e.~it has no lattice artefacts on field configurations satisfying
the equations of motions. For each gauge field configuration on the
lattice, the minimizing gauge field $U(x,V)$ defines a FP field in the
continuum which induces some remarkable properties on the lattice
gauge fields. Under certain conditions, all symmetries of the
continuum are well defined on the lattice \cite{Hasenfratz:2006kv} and
the FP field allows for a representation of the corresponding
infinitesimal transformations on the lattice according to
\[
V_n \quad \stackrel{\tiny\text{minimize}}{\longrightarrow} \quad
U(x,V) \quad \stackrel{\tiny\text{transform}}\longrightarrow \quad
U^\varepsilon(x^\varepsilon,V) \quad
\stackrel{\tiny\text{block}}\longrightarrow \quad V^\varepsilon_n \, .
\]
Similar considerations can be made for the fermion fields
\cite{Hasenfratz:2006kv}.  In particular, the procedure in principle
also provides a possibility to define supersymmetric algebras on the
lattice. The fact that the FP equations define a scheme to match
coarse and fine lattice configurations could also be useful in recent
attempts to delay the onset of topological critical slowing down in
today's lattice QCD simulations by constructing multi-scale
Monte-Carlo update algorithms. One effort in this direction
\cite{Endres:2015yca,Detmold:2016rnh}, in which the FP action approach
could be very useful, is in fact discussed in a plenary talk at this
conference \cite{Endres:2016rzj}.

\begin{figure}[tbh]
\centering
\includegraphics[width=0.8\textwidth]{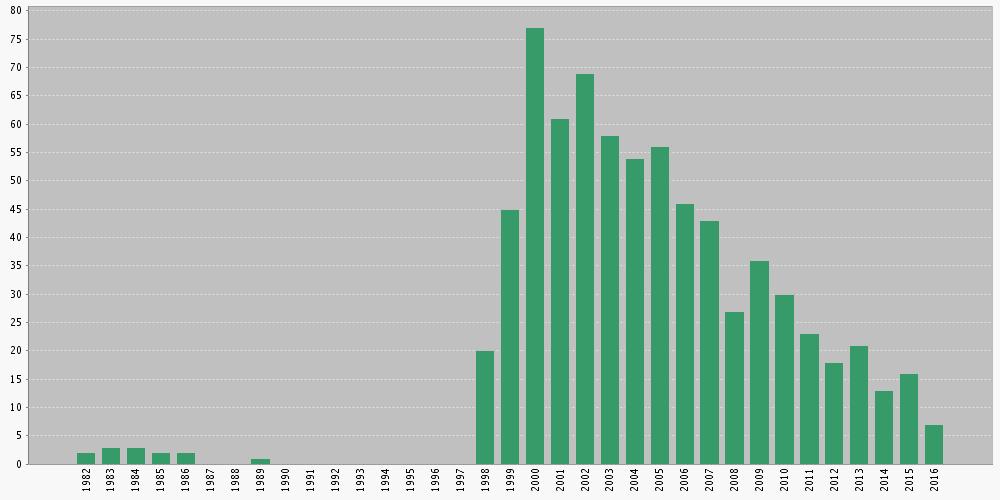}
\caption{Citation history of the  Ginsparg-Wilson paper \cite{Ginsparg:1981bj}.\label{fig:citationHistory_GW}}
\end{figure}

Then, in 1997 Peter made a truly groundbreaking observation
\cite{Hasenfratz:1997ft}. While travelling to a summer school Peter
looked through a pile of old preprints which he picked up while
tidying up
\begin{wrapfigure}{r}{0.7\textwidth}
\includegraphics[width=0.7\textwidth]{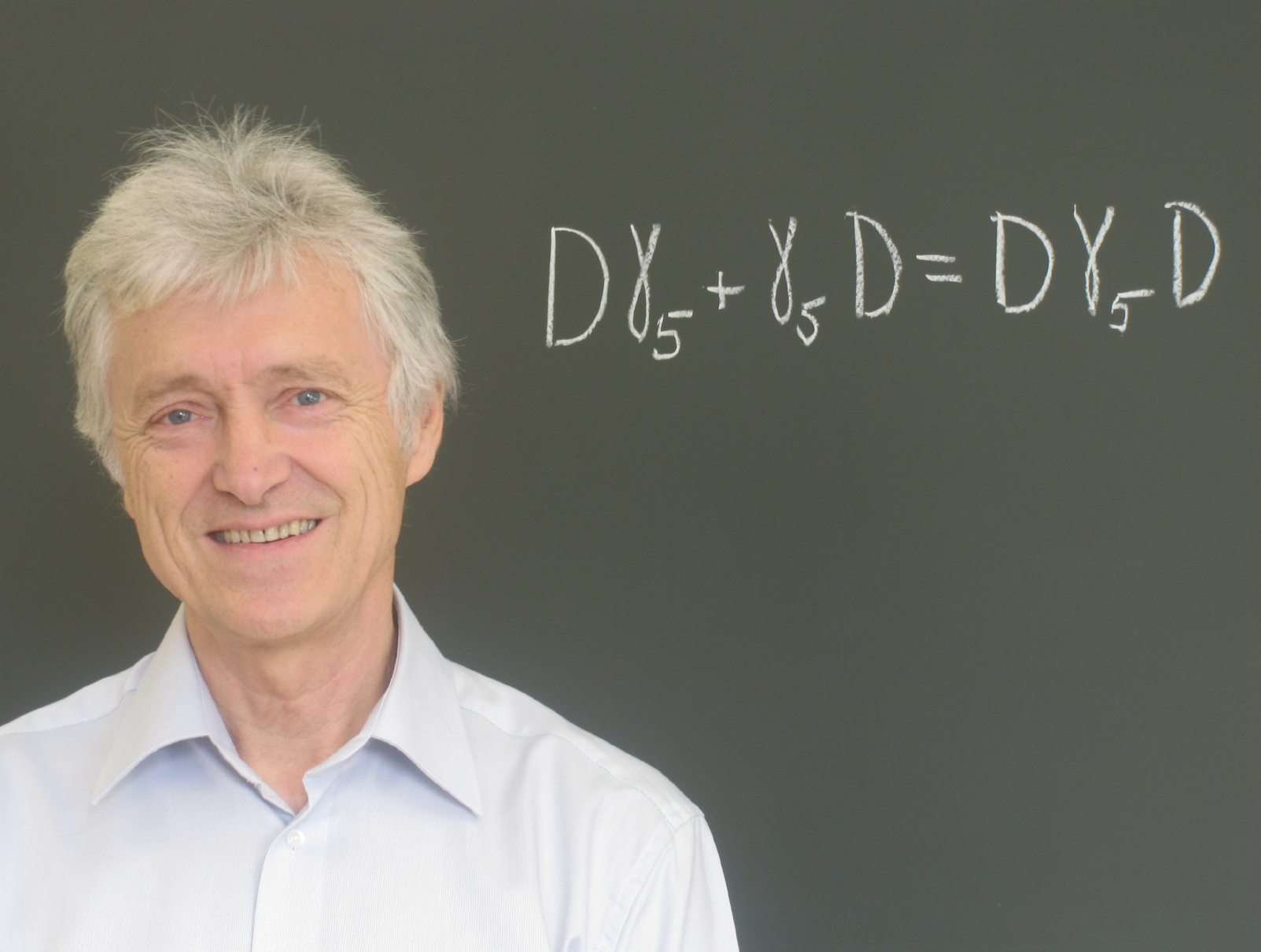}
\end{wrapfigure}
his office. A paper by Ginsparg and Wilson \cite{Ginsparg:1981bj}
grabbed his attention and he realized that the FP Dirac operator
$D^\text{FP}$ fulfills the now-famous Ginsparg-Wilson relation
\[
D \gamma_5 + \gamma_5 D = D \gamma_5 D \, .
\]
The relation is derived from RG transformations applied to free
fermions. Any solution of the relation avoids the Nielsen-Ninomyia
no-go theorem, implies the correct triangle anomaly and the validity
of all the soft-pion theorems on the lattice
\cite{Hasenfratz:1997ft}. Peter's crucial observation then was that
the FP Dirac operator $D^\text{FP}$ constitutes a solution for the
interacting theory. As a consequence of the Ginsparg-Wilson relation
there is no tuning, no mixing and no current renormalization necessary
for the FP Dirac operator on the lattice~\cite{Hasenfratz:1998jp}. The
observation set off an avalanche of developments. This is best
visualized by looking at the citation history of the Ginsparg-Wilson
paper \cite{Ginsparg:1981bj},
cf.~Fig.~\ref{fig:citationHistory_GW}. It is fascinating to see how
the interest in the paper exploded after Peter's rediscovery in 1997
-- with 985 citations it is by now the second most cited paper cited
by the hep-lat archive \cite{INSPIRE:topcite2016}. It is impossible to
describe in detail the revolution which followed, so let me just
mention a few key developments on the theoretical side: the exact
index theorem in QCD on the lattice \cite{Hasenfratz:1998ri}; the
overlap operator as another solution to the Ginsparg-Wilson relation
\cite{Neuberger:1998wv}; the exact chiral symmetry on the lattice
\cite{Luscher:1998pqa}; Abelian chiral gauge theories on the lattice
\cite{Luscher:1998du, Neuberger:1998xn}; the axial anomaly and
topology
\cite{Kikukawa:1998pd,Luscher:1998kn,Chiu:1998xf,Adams:1998eg}; the
chiral Jacobian on the lattice \cite{Fujikawa:1998if,Suzuki:1998yz};
lattice supersymmetry \cite{So:1998ya}; and so on and on.

Even today, the implications for phenomenological and theoretical
applications can hardly be overestimated, and as evidence I just refer
to the various parallel sessions such as "Weak decays and matrix
elements", "Chiral symmetry", "Theoretical developments", with several
talks related to or based on chiral symmetry on the lattice in some
way or another.

In order to close this section, let me make a couple of remarks which
I find particularly intriguing in the context of chiral symmetry
regularized on the lattice. Firstly, the possibility to realize
exactly massless fermions on the lattice constitutes a true solution
of a hierarchy problem, a fact which is often unappreciated outside
the lattice community. Secondly, FP fermions on the one hand and
overlap/domain wall (DW) fermions
\cite{Kaplan:1992bt,Narayanan:1992wx,Shamir:1993zy,Neuberger:1997fp}
on the other hand derive from two completely different approaches, yet
they both fulfill the Ginsparg-Wilson relation.  Is there a connection
between the two formulations, and if yes, how are they related?  While
the FP operator is motivated from RG considerations and therefore has
a very specific physical meaning, the overlap/DW operator has a less
obvious physical interpretation. Rather, the overlap/DW operator can
more easily be understood from an algebraic viewpoint. More precisely,
the Ginsparg-Wilson relation is a specific form of an {\it algebraic
  Riccati equation}, the solution of which naturally involves the sign
function as it also appears in the overlap/DW operator. Thirdly, the
chiral transformation on the lattice makes use of the gauge field
dependent chiral projectors $\hat P_\pm = 1/2(1\pm\hat\gamma_5)$ with
$\hat \gamma_5 = \gamma_5(1-D)$. They are particularly important in
the context of defining chiral gauge theories on the lattice, a
problem which remains to be of high interest and is in fact also
discussed in a plenary talk at this lattice conference,
cf.~\cite{Grabowska:2016bis} and references therein to the
corresponding lattice talks. On the one hand, the chiral projectors
are responsible for the necessary asymmetry between fermions and
anti-fermions, on the other hand they also break $CP$ symmetry
\cite{Hasenfratz:2007dp}. The $CP$ breaking is an unwanted feature,
the role of which is still not well understood.

\section{Closing remarks}
After my recollection of just three out of many of Peter's scientific
contributions, it is important to realize that Peter's legacy is much
more than his outstanding scientific achievements. It goes far beyond
what he has taught us about quantum field theories on and off the
lattice.
As researchers we are all curious and strive for the unknown, but
Peter taught us what it means to go one step further: do look close
and careful, be {\it very} critical - and do question even the
presumably well-known and the commonly accepted. This will inevitably
lead to surprising results, new insights and creative ideas. Of course
this is all said easily and quickly - it distinguishes Peter that he
was indeed able to follow this path and operate in such a perfect way.
The careful and conservative path is usuallly not very spectacular,
but it is certainly much more enduring and lasting, and, as Peter
pointed out, worth following. It will sustain and support the lattice
community much better in the long term by gaining and keeping a high
regard and esteem from outside our community.

One of Peter's legacy obviously is the fact that we all meet once a
year to discuss our progress, exchange the latest insights and
results, and get excited about new developments. I am very happy to
see that this legacy is carried on.  Looking into the audience I see
many young faces among the very many participants, so I am very
optimistic that this legacy will continue for a very long time -- I am
certain that this would make Peter very happy.

I would like to thank Anna Hasenfratz, Kieran Holland, Ferenc
Niedermayer and Uwe-Jens Wiese for helpful discussions during the
preparation of this talk.

\bibliographystyle{JHEP}
\bibliography{fsmtlQCDtsloPH}

\end{document}